\documentclass[showpacs,preprintnumbers,amsmath,amssymb,aps,twocolumn,pra]{revtex4-1}

\usepackage{graphicx}
\usepackage{bm}

\renewcommand\vec[1]{\mathbf{#1}}

\begin{document}
\title{
Controlling and Detecting Spin Correlations of Ultracold Atoms in Optical Lattices
}

\author{Stefan Trotzky$^{1-3}$}
\author{Yu-Ao Chen$^{1-3}$}
\author{Ute Schnorrberger$^{1-3}$}
\author{Patrick Cheinet$^{4}$}
\author{Immanuel Bloch$^{1-3}$}

\affiliation{$^1$ Fakult\"at f\"ur Physik, Ludwig-Maximilian-Universit\"at, Schellingstrasse 4, 80798 M\"unchen, Germany}
\affiliation{$^2$ Max-Planck-Institut f\"ur Quantenoptik, Hans-Kopfermann-Strasse 1, 85748 Garching, Germany }
\affiliation{$^3$ Institut f\"ur Physik, Johannes Gutenberg-Universit\"at, Staudingerweg 7, 54099 Mainz, Germany}
\affiliation{$^4$ Laboratoire Charles Fabry, Institut d'Optique, Campus Polytechnique, RD 128, 91127 Palaiseau Cedex, France}

\date{\today}

\pacs{03.67.Pp, 67.85.Jh, 75.10.Kt}

\begin{abstract}
We report on the controlled creation of a valence bond state of delocalized effective-spin singlet and triplet dimers by means of a bichromatic optical superlattice. We demonstrate a coherent coupling between the singlet and triplet states and show how the superlattice can be employed to measure the singlet-fraction employing a spin blockade effect. Our method provides a reliable way to detect and control nearest-neighbor spin correlations in many-body systems of ultracold atoms. Being able to measure these correlations is an important ingredient to study quantum magnetism in optical lattices. We furthermore employ a SWAP operation between atoms being part of different triplets, thus effectively increasing their bond-length. Such SWAP operation provides an important step towards the massively parallel creation of a multi-particle entangled state in the lattice.
\end{abstract}

\maketitle

Strong correlations in quantum many-body systems are a cornerstone of modern condensed-matter physics. They underlie the Mott insulator (MI) state of electrons in the cuprates, which feature an extremely rich phase diagram \cite{Sachdev:2003} and exhibit high-$T_c$ superconductivity upon doping \cite{Bednorz:1986,*Lee:2006}. The experimental realization of the MI with ultracold bosonic \cite{Greiner:2002a,*Stoferle:2004,*Spielman:2007} and -- more recently -- fermionic atoms \cite{Joerdens:2008,*Schneider:2008} in optical lattices has demonstrated the prospect of these system to address fundamental condensed-matter problems \cite{Bloch:2008,*Lewenstein:2007}. One crucial requirement for the study of quantum magnetism with ultracold atoms is a sensitive probe of spin correlations that characterize magnetic phases and can be employed to determine the entropy of the system \cite{Joerdens:2009}. Furthermore, the controlled manipulation of nearest-neighbor spin correlations might enable to create low-entropy spin-correlated states that can e.g. be adiabaticaly connected to coupled dimer states or an antiferromagnetically ordered state, circumventing cooling problems \cite{GarciaRipoll:2004,*Sorensen:2010}.

In this paper, we demonstrate both the control and the detection of short-range spin-correlations with ultracold bosons in an optical superlattice. We create a three-dimensional array of effective-spin triplet pairs and induce coherent singlet-triplet oscillations (STO) on the bonds. We make use of the different parity of the singlet and triplet wavefunctions to distinguish the two after merging pairs of sites \cite{Paredes:2008}. The underlying mechanism is the singlet-triplet blockade known from double-electron quantum dots \cite{Petta:2005,*Johnson:2005,*Hanson:2007}. The detection procedure can be applied to directly measure the amount of nearest-neighbor singlet and triplet correlations in a two-species MI of neutral atoms making it a valuable method to measure the entropy, as for example the singlet density in a fermionic MI at half filling is expected to increase with decreasing temperatures \cite{Paiva:2001,*Fuchs:2010}. 
We furthermore employ a massively parallel SWAP gate between neighboring triplet pairs which provides an important step towards the generation of multi-particle entangled states with possible application in measurement based quantum computation \cite{Raussendorf:2001,*Raussendorf:2003}.

Our experimental setup consists of a three dimensional optical superlattice filled with $^{87}$Rb atoms \cite{Foelling:2007a}. 
Along the $x$-direction, the superlattice is formed by two collinear retro-reflected laser beams of wavelengths $\lambda_{xs} = 765\,{\rm nm}$ (short lattice) and $\lambda_{xl} = 1530\,{\rm nm}$ (long lattice). The relative phase $\phi$ of the two standing waves can be adjusted freely to yield a potential of the form $V(x) = V_{xs} \cos(4 k_x x) - V_{xl} \cos(2 k_x x+\phi)$ with $k_x = 2\pi / \lambda_{xl}$ and where the lattice depths $V_{xs}$ and $V_{xl}$ are controllable by the intensity of the individual laser beams (see Fig.~1a). Transverse monochromatic lattices with wavelengths $\lambda_{y,z}=843\,{\rm nm}$ complete the three-dimensional array of double-wells (DWs). Furthermore, a magnetic field gradient $\partial_x B_x$ can be applied to create a potential bias in the DWs which depends on the Zeeman state of the atoms. 

\begin{figure}[t]
\includegraphics[scale=0.8]{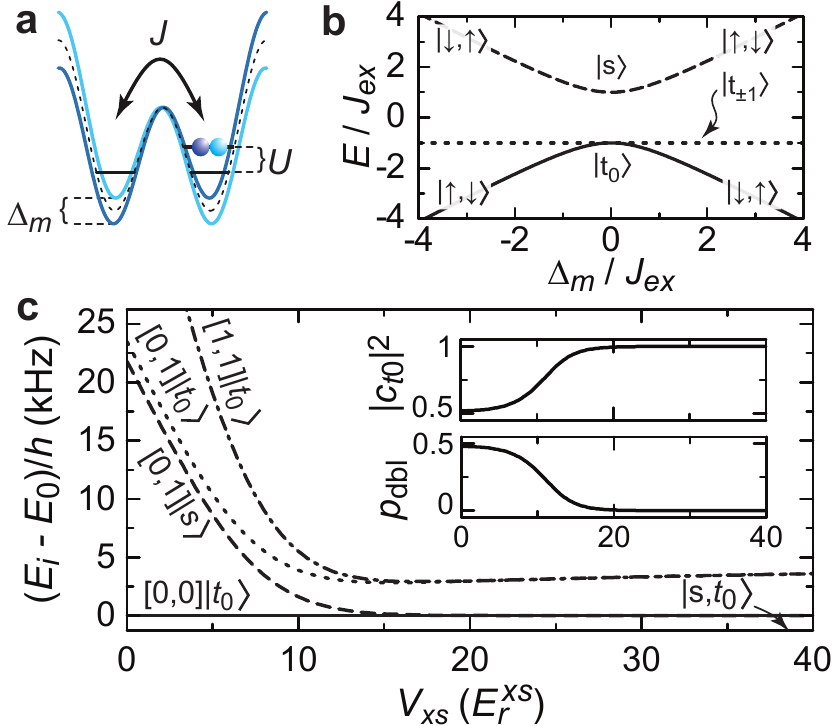}
\caption{\textbf{a} Schematic drawing of the DWs formed by the superlattice.
\textbf{b} Eigenstates of Eq.~(2) for two particles versus $\Delta_m$. At zero gradient, the states $|\!\!\uparrow,\downarrow\rangle$ and $|\!\!\downarrow,\uparrow\rangle$ couple to yield the triplet state $|t_0\rangle$ and the singlet $|s\rangle$. \textbf{c} Relative eigenenergies for two $^{87}$Rb atoms in the $|F=1,m_F=\pm1\rangle$ Zeeman states in the DW potential versus barrier height $V_{xs}$. The notation $[v_1,v_2]$ for $V_{xs} \to 0$ refers to the vibrational quantum numbers for the first and second particle. The insets show the overlap $|c_{t0}|^2$ of the groundstate with the triplet $|t_0\rangle$ and the amount of double-occupancy $p_{dbl}$.\vspace{-3ex}}
\end{figure}

We now consider the situation of symmetric DWs ($\phi = 0$), occupied by two atoms in two internal states labeled $|\!\!\uparrow\rangle$ and $|\!\!\downarrow\rangle$. If the largest energy scale in the problem is given by the trap frequencies of the individual wells, the system can be described by a two-site Bose-Hubbard type model
\begin{eqnarray}
      \hat H
  &=& - J \sum_{\sigma=\uparrow,\downarrow} \left(\hat a_{\sigma L}^{\dagger} \hat a_{\sigma R}^{\phantom{\dagger}} + {\rm h.c.}\right) \nonumber\\
  &&  + U/2 \left(\hat n_{L} (\hat n_{L} - 1) + \hat n_{R} (\hat n_{R} - 1) \right) \nonumber\\
  &&  + \Delta_m/2 \left(\hat n_{\uparrow L}-\hat n_{\downarrow L}-\hat n_{\uparrow R}+\hat n_{\downarrow R}\right)\,,
\end{eqnarray}
with the tunnel coupling $J$, the on-site interaction energy $U$ (see Fig.~1a) and the state-dependent bias of magnitude $\Delta_m$ which reflects the magnetic field gradient. The operator $\hat a_{\sigma L(R)}$ annihilates a particle in the spin state $\sigma$ localized in the left (right) well, $\hat n_{\sigma,L(R)} = \hat a^\dagger_{\sigma L(R)} \hat a^{\phantom{\dagger}}_{\sigma L(R)}$ counts the number of particles per state and well and $\hat n_{L(R)} = \sum_\sigma \hat n_{\sigma L(R)}$.

For strong repulsive interactions ($U \gg J$), the groundstate manifold is characterized by an occupation of one particle per site \cite{Sebby:2007a}. Virtual tunneling processes via higher-energy states mediate an effective superexchange coupling in this subspace \cite{Auerbachbook,Anderson:1950}. The corresponding effective Hamiltonian can be derived from Eq.~(1) \cite{Duan:2003,*Garcia-Ripoll:2003,*Kuklov:2003}:
\begin{eqnarray}
  \hat H_{\rm eff} &=& - J_{ex} (\hat{ \vec S}_L \cdot \hat{\vec S}_R + \hat{\vec S}_R \cdot \hat{\vec S}_L)
+ \Delta_m (\hat S_L^z - \hat S_R^z)\,,
\end{eqnarray}
with the (ferromagnetic) superexchange coupling $J_{ex} = 2J^2/U > 0$ and the effective spin-1/2 operators $S_i^{x} = (\hat a_{\uparrow i}^\dagger \hat a^{\phantom{\dagger}}_{\downarrow i} + \hat a_{\downarrow i}^\dagger \hat a^{\phantom{\dagger}}_{\uparrow i})/2$, $S_i^{y} = (\hat a_{\uparrow i}^\dagger \hat a^{\phantom{\dagger}}_{\downarrow i} - \hat a_{\downarrow i}^\dagger \hat a^{\phantom{\dagger}}_{\uparrow i})/2i$ and $\hat S_i^z = (\hat n_{\uparrow i} - \hat n_{\downarrow i})/2$. This simple two-site model is well known within the framework of electrons in double-dots \cite{Petta:2005, Hanson:2007}. The groundstate of the Hamiltonian Eq.~(2) for $J_{ex}>0$ and $\Delta_m = 0$ is the three-fold degenerate effective-spin triplet which consists of $|\rm t_{-1}\rangle = |\!\!\downarrow,\downarrow\rangle$, $|\rm t_{+1}\rangle = |\!\!\uparrow,\uparrow\rangle$ and $|\rm t_0\rangle = (|\!\!\uparrow,\downarrow\rangle+|\!\!\uparrow,\downarrow\rangle)/\sqrt{2}$. The singlet state $|\rm s\rangle = (|\!\!\uparrow,\downarrow\rangle-|\!\!\uparrow,\downarrow\rangle)/\sqrt{2}$ is higher in energy by $2J_ {ex}$ (see Fig.~1b). For $\Delta_m \gg J_{ex}$, the degeneracy of the states $|\!\!\uparrow,\downarrow\rangle$ and $|\!\!\downarrow,\uparrow\rangle$ is lifted and they become the eigenstates replacing $|s\rangle$ and $|t_0\rangle$. A coherent coupling of $|s\rangle$ and $|t_0\rangle$ is realized by rapidly changing $\Delta_m$ from zero to $\Delta_m \gg J_ {ex}$, projecting onto these new eigenstates. The subsequent phase-evolution describes singlet-triplet oscillations with a frequency $\nu_{STO} \simeq 2 \Delta_m/h$. Previously, the reverse situation of a coherent superexchange coupling of the states $|\uparrow,\downarrow\rangle$ and $|\downarrow,\uparrow\rangle$ at $\Delta_m=0$ has been realized in the same system \cite{Trotzky:2008}.

When the barrier is removed adiabatically ($V_{xs} \to 0$), the singlet and triplet states in the DW are transferred into the two-particle eigenstates of the underlying long-lattice well. In Fig.~1c, we plot the eigenenergies $E_i$ of the two particles in the DW with respect to the groundstate energy $E_0$ as a function of $V_{xs}$ as obtained from an extended two-site Hubbard model \cite{SuppMat}. As the barrier is ramped down, the spin-symmetry of the state is preserved.
While the triplet states are adiabatically connected to the two-particle vibrational groundstate $[0,0]|t_0\rangle$ of the underlying long-lattice well, the singlet state requires one particle to occupy the first excited orbital, thus $[0,1]|s\rangle$. This can be seen as an analog to the spin-blockade in electronic quantum dots \cite{Johnson:2005}. To fulfill the bosonic statistics, the spatial symmetry of the two-body wavefunction has to match the spin-symmetry \cite{FermionNote1}.
In consequence, merging pairs of sites and subsequently measuring the number of band excitations can be used to distinguish triplet and singlet correlations in a two-species MI. The on-site exchange splitting between $[0,1]|s\rangle$ and $[0,1]|t_0\rangle$ could previously be measured via on-site exchange oscillations \cite{Anderlini:2007}. The ability to coherently couple $|t_0\rangle$ to $|s\rangle$ by means of the gradient term $\Delta_m$ further provides the possibility to distinguish the states $|\!\!\uparrow,\downarrow\rangle$ and $|\!\!\downarrow,\uparrow\rangle$ from $|t_0\rangle$ and $|s\rangle$.

We begin our experiments by loading a BEC of about $9\times 10^4$ $^{87}$Rb-atoms in the $|F=1,m_F=-1\rangle$ Zeeman state from a magnetic trap with high offset field into the 3D optical lattice with $V_{xs} = 36\,E_r^{xs}$ and $V_{y,z}=35\,E_r^{y,z}$ \cite{ErecoilNote}. The resulting state is a MI with at most one atom per site. By raising the long lattice to $V_{xl} = 40\,E_r^{xl}$ and removing the short lattice, we merge pairs of lattice sites to yield constructed atom pairs in the vibrational groundstate in the long-lattice wells \cite{SuppMat}. We use a radio-frequency rapid adiabatic passage (rfRAP) to transfer the atoms into the $|1,0\rangle$ Zeeman state and afterwards switch off the magnetic trap, maintaining a homogeneous offset field of $\simeq 1.2\,{\rm G}$. Atom pairs in $|1,0;1,0\rangle$ are subsequently transferred into $|1,+1;1,-1\rangle$ triplet pairs by means of spin-changing collisions (SCC) \cite{Widera:2005}. The two Zeeman states $|1,+1\rangle$ and $|1,-1\rangle$ serve as the two effective spin states $|\!\!\uparrow\rangle$ and $|\!\!\downarrow\rangle$ \cite{Trotzky:2008}.

We split the long-lattice wells into symmetric DWs by ramping up the short lattice to $40\,{E_r^{xs}}$ again within $10\,{\rm ms}$ which creates an array of delocalized triplets $|t_0\rangle$. When the superexchange coupling is fully suppressed, we switch on a magnetic field gradient of variable strength for a holdtime $t_{\rm hold}$. We detect the emerging STO by ramping down the barrier and employing a band-mapping technique \cite{Greiner:2001b}. A pulsed magnetic field gradient at the beginning of the expansion separates the different Zeeman states. In order to avoid spurious signals due to imperfections of the rfRAP and the SCC, the band-excitations are measured for the $|F=1,m_F=+1\rangle$ Zeeman state alone -- the only one not present before the SCC. The loading of the array of triplets and the subsequent conversion into singlets yields the analogue of a valence bond solid (VBS) state with bosons \cite{Sachdev:2003,Lhuillier:2002,Sachdev:2008}. This state is characterized by a bond order along the superlattice direction with the period of the long lattice.

In Fig.~2a, we plot the measured fraction of band excitations after merging $n_{\rm exc}(t_{\rm hold})$ for the $|F=1,m_F=1\rangle$ Zeeman state for a gradient of about $10\,{\rm G/cm}$. We observe an amplitude of the STO of about 30\%. The reduction from the ideal value of 50\% stems from a residual magnetic field gradient present during the splitting, the finite lifetime of the triplet and singlet state and imperfections of the detection method. About 4\% of band-excitations created during the loading procedure give rise to an additional offset. The phase-shift of the oscillations is a result of the finite switch-on time of the gradient. Fig.~2b shows the measured oscillation frequency $\nu_{STO}$ which depends linearly on the current applied to the coil producing the gradient. We independently measure the energy bias $h\nu_{SP}$  between the left and the right-well for a single particle versus gradient strength. This is done by coherently splitting long-lattice sites with single atoms in $|\!\!\downarrow\rangle$ and recording the emerging interference patterns after a short holdtime $\simeq 1\,{\rm ms}$ and time-of-flight \cite{Sebby:2007a}. We find the STO frequency to be $2\nu_{SP}$ (see Fig.~2b) which confirms the two-body nature of the measured effect. From a linear fit of $\nu_{\rm STO}$ we find a gradient offset of $\simeq 120\,{\rm mG/cm}$ which might stem from inhomogeneities of the $1.2\,{\rm G}$ offset field and static magnetic field sources close to the experiments. An alternative explanation is a relative angle between the polarizations of the two lattice beams of the short lattice of $<1\,{\rm mrad}$ leading to a weak spin-dependency of the superlattice \cite{Deutsch:1998}.

\begin{figure}[t]
\includegraphics[scale=0.8]{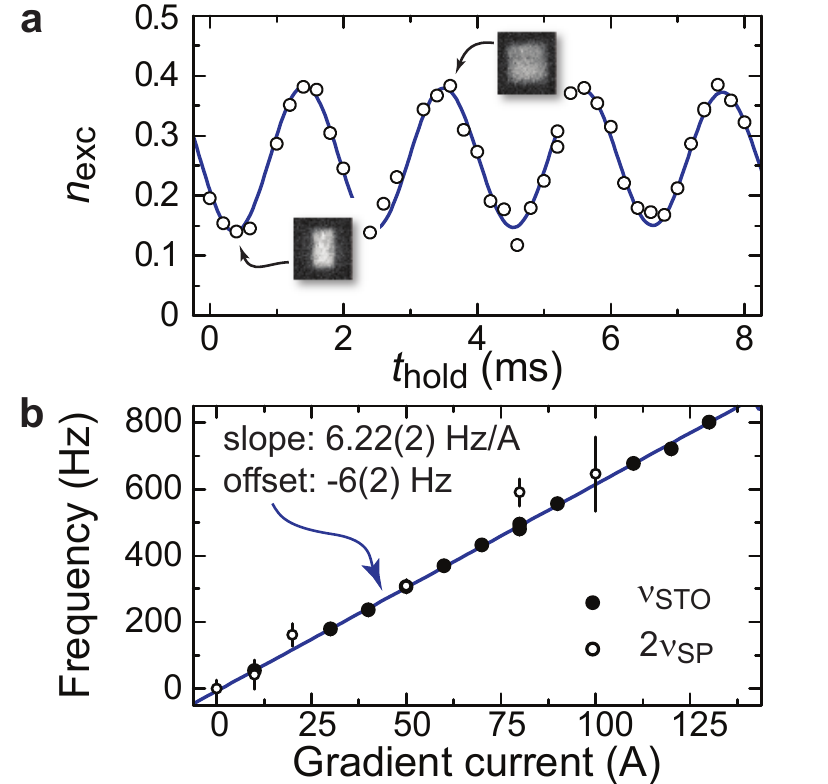}
\caption{\textbf{a} Plot of the relative population in the vibrationally excited band $n_{\rm exc}(t_{\rm hold})$ for a gradient current of $80\,{\rm A}$ (circles) together with the fit of a sine-wave (solid line). The insets show typical absorption images obtained by the detection sequence. \textbf{b} Fitted oscillation frequencies $\nu_{STO}$ versus gradient current (filled circles) together with an independent measurement of the single-particle shift $\nu_{SP} = \Delta_m/h$ (open circles).
The solid line is a linear fit to $\nu_{STO}$ which translates into a slope of $0.122(1)\,{\rm G / (cm\,A)}$ and an offset of $-0.12(3)\,{\rm G/cm}$.\vspace{-3ex}}
\end{figure}

For all gradient strengths, we find a damping time of the oscillations of about $\tau = 40-50\,{\rm ms}$. This damping is mainly caused by the decay of the triplet and the singlet state. We measure the lifetime of the triplet by merging the DWs after the holdtime $t_{\rm hold}$ without applying the magnetic field gradient. The same is done for the singlet state, where only a short gradient pulse after the splitting is employed to convert $|t_0\rangle$ into $|s\rangle$. In Fig.~3, we plot $n_{exc}$ for the two measurements as a function of $t_{\rm hold}$ together with a STO trace taken at a gradient current of $80\,{\rm A}$. The lifetime measurement provides an envelope to the oscillation data. A simultaneous fit of the triplet- and singlet-lifetime yields $\tau = 43(1)\,{\rm ms}$ which matches the damping-time of the STO. This lifetime gradually decreases when increasing the short-lattice and transverse-lattice depths. Neither spontaneous scattering of lattice photons ($\Gamma_{sc}^{-1} > 500\,{\rm ms}$), nor  tunneling of individual particles can explain this lifetime. It is most likely limited by weak spin dependencies of the external confinement due to imperfections in the polarization of the lattice beams \cite{SuppMat}. A larger detuning of the lattice beams would significantly reduce the sensitivity on these imperfections. 

\begin{figure}[t]
\includegraphics[scale=0.8]{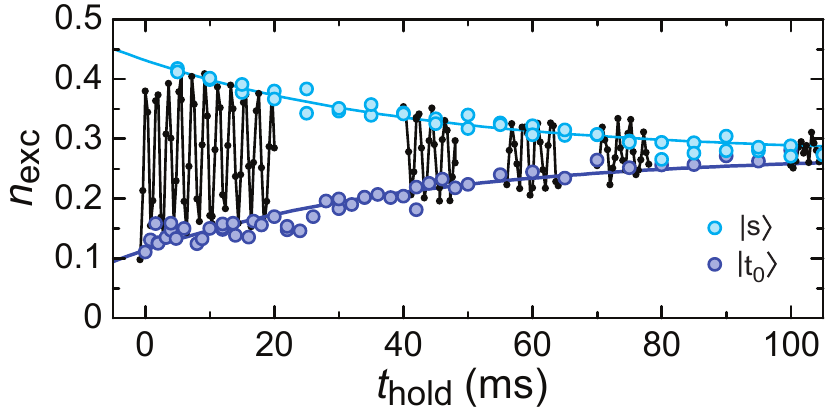}
\caption{Measured lifetime of the triplet (dark blue) and the singlet state (light blue) versus holdtime in the split DWs. The blue solid lines are the result of a simultaneous exponential fit yielding a lifetime of $43(1)\,{\rm ms}$. The black data points show STO taken with the same sequence and an additional gradient field of about $10\,{\rm G/cm}$ during the holdtime.\vspace{-3ex}}
\end{figure}

In addition to the formation and detection of the delocalized triplets, the superlattice also offers unique possibilities for further manipulation (Fig.~4a), i.e. aimed at generating multi-particle entangled states or to increase the spatial extension of the entangled spin pairs \cite{Barmettler:2008}. After having created the array of $|t_0\rangle$ bonds on neighboring sites, we remove the long-lattice, shift its phase to $\phi = \pi$ and ramp it high again, thus combining pairs of sites which belong to different triplet-bonds to a DW. By lowering the short lattice to $12\,E_r^{xs}$ within $200\,{\rm\mu s}$, we switch on a superexchange coupling between these sites with $J_{ex} = J/3 = h\times 360\,{\rm Hz}$ \cite{Trotzky:2008}. We ramp the barrier high again after half a superexchange period which realizes a SWAP operation for the coupled spins. As a result, the triplets are now delocalized over three lattice spacings rather than a single one before \cite{Barmettler:2008}. We again use the magnetic field gradient to induce the STO and subsequently repeat the SWAP operation, in order to restore the original bond length of the entangled spin pairs. Finally, the same combination of merging and band mapping is used as before to reveal the STO. The measured fraction $n_{exc}(t_{\rm hold})$ is plotted in Fig.~4b together with a trace recorded with the same sequence but without lowering the short lattice to induce the two SWAPs. We find an oscillation frequency three-times higher when the SWAP operations are carried out. This is explained by the linear dependence of $\Delta_m$ on the distance of the particles for a given gradient strength. The corresponding STO amplitude is reduced by about $40\%$ which in parts stems from unsuccessful SWAPs where there is a hole adjacent to a triplet bond. These holes are detected as additional excitations, yielding a higher offset of the STO.

\begin{figure}[t]
\includegraphics[scale=0.8]{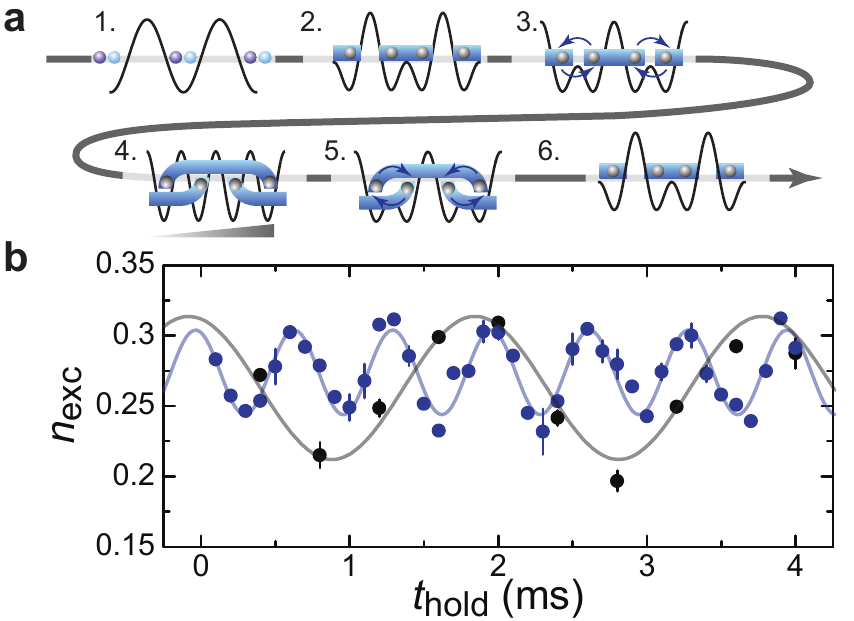}
\caption{\textbf{a} Scheme of the SWAP-sequence: The spin pairs (1) are split into triplets (2). A superexchange for half a period SWAPs the spins of neighboring triplets (3) before a magnetic field gradient is used to drive STO (4). A second SWAP (5) brings the spins back to their original position (6). \textbf{b} Measured excited fraction after the final merging with (blue circles) and without (black circles) applying the SWAP gates. The oscillations with the SWAP are faster by a factor of 2.91(10), confirming the successful stretching of the spin-pairs.\vspace{-3ex}}
\end{figure}

The successful demonstration of the SWAP operation constitutes an important step to further entangle neighboring triplet pairs which can be achieved by a $\sqrt{\rm SWAP}$ operation. Since this manipulation is carried out between all neighboring singlets simultaneously, the result after a single step would be an entangled chain of neutral atoms. With a second superlattice along a perpendicular direction, this entanglement can be extended on a 2D plane in the same manor. The resulting state is maximally entangled in 2D and therefore would be valuable for measurement based quantum computation.

In conclusion, we have demonstrated the controlled loading of an array of nearest-neighbor $S_z = 0$ triplets and the conversion into singlets by means of an optical superlattice and a magnetic field gradient. The array of singlet or triplet dimers can be seen as a valence bond solid with the bonds alternating along the $x$-axis. We note that the same method can be applied to create a coupled dimer antiferromagnet of fermions \cite{Sachdev:2008} from an initial low-entropy band-insulating state. We have also established a sensitive method to measure nearest-neighbor spin correlations to reveal gradient driven singlet-triplet oscillations. Finally we have demonstrated the controlled stretching of the triplet pairs by a massively parallel superexchange SWAP gate between neighboring bonds. Bringing distant spins together in a DW by SWAP operations is a way to measure longer-range spin-correlations. The implementation of a $\sqrt{\rm SWAP}$ gate instead can be used to entangle chains -- and further 2D arrays -- of neutral atoms.

We are grateful to Simon F\"olling, Artur Widera and Michael Feld for important help in the early stage of this experiment. We acknowledge stimulating discussion with Bel\'en Paredes and Jens Eisert.
This project was funded by the DFG (FOR635), DARPA (OLE-Program) and the EU (NAMEQUAM).

\section*{Appendix I -- Extended Bose-Hubbard model}
We calculate the eigenenergies $E_i$ in the double-wells as a function of $V_{xs}$ (Fig.~1a) using an extension of the Bose-Hubbard model Eq.~(1). Since the overlap of the left- and right-localized wavefunctions becomes significant for low barriers -- thus small values of $V_{xs}$ -- a direct nearest-neighbor interaction needs to be taken into account, as well as a density-dependent modification of the tunnel coupling \cite{Trotzky:2008}. The corresponding extended Bose-Hubbard Hamiltonian reads
\begin{eqnarray*}\label{eq:EBHM}
 \hat  H_{\rm EBHM} &=& \hat H - \Delta J \sum_{\sigma \neq \sigma'} (\hat n_{\sigma L} + \hat n_{\sigma R})
  \left(\hat a_{\sigma'L}^\dagger\hat a_{\sigma'R}^{\phantom{x}} + \text{h.c.}\right) \nonumber\\
&& + U_{LR} \sum_{\sigma \neq \sigma'} \Big( \hat n_{\sigma L}\hat n_{\sigma' R} 
  + \hat a_{\sigma L}^\dagger\hat a_{\sigma' R}^\dagger \hat a_{\sigma' L}^{\phantom{x}}\hat a_{\sigma R}^{\phantom{x}} \nonumber\\
&& \phantom{+ U_{LR} \sum_{\sigma \neq \sigma'}} + \frac{1}{2}\,
  \hat a_{\sigma L}^\dagger\hat a_{\sigma' L}^\dagger \hat a_{\sigma' R}^{\phantom{x}}\hat a_{\sigma R}^{\phantom{x}} \nonumber\\
&& \phantom{+ U_{LR} \sum_{\sigma \neq \sigma'}} + \frac{1}{2}\,
  \hat a_{\sigma R}^\dagger\hat a_{\sigma' R}^\dagger \hat a_{\sigma' L}^{\phantom{x}}\hat a_{\sigma L}^{\phantom{x}} \Big)\,,
\end{eqnarray*}
with $U_{LR} = g \times \int w_L^2(\vec r) w_R^2(\vec r)\, d\vec r$ and $\Delta J = -g \times\int w_L^3(\vec r) w_R(\vec r)\,d\vec r$. The functions $w_{L,R}(\vec r)$ are the left- and right-localized groundstate wavefunctions functions in the double-well. The prefactor to the integrals is $g = 4\pi^2 \hbar^2 a_S / m_{\rm Rb}$ with $a_S \simeq 5.3\,{\rm nm}$ being the $s$-wave scattering length and $m_{\rm Rb}$ the mass of the $^{87}$Rb atoms. All parameters of the extendend Bose-Hubbard model Eq.~(3) are obtained from a single-particle bandstructure calculation.

\section*{Appendix II -- Loading of atom pairs in $|F=1, m_F=0\rangle$}
During the rampup of the lattice 3D optical lattice to load the MI, we apply a shallow long-lattice potential of $V_{xl} = 1\,E_r^{xl}$ with a phase $\phi \simeq 0.2$. Therefore, possible defects are loaded mostly to ``left" sides of the double-wells. When merging the pairs of atoms, we apply a small bias $\Delta < U$ ($V_{xl} = 40\,E_r^{xl}$, $\phi \simeq 0.05$), to avoid band-excitations from these singly occupied double-wells. As a result of the loading sequence, we obtain $\simeq 60\%$ of the atoms loaded into pairs.

The offset-field of the magnetic trap from which the optical lattice is loaded is about $153\,{\rm G}$. After loading the MI, we transfer all atoms from $|F=1, m_F=- 1\rangle$ to $|1,0\rangle$ by coupling them with a radio-frequency signal of $108.6\,{\rm MHz}$ and sweeping the magnetic field across the resonance at $\simeq 152.5\,{\rm G}$. We optimize the magnetic field ramp to obtain no population in $|1,1\rangle$ and achieve a transfer efficiency of more than $96\%$. Due to the quadratic Zeeman shift, the transition $|1,0\rangle \leftrightarrow |1,1\rangle$ is detuned by about $3\,{\rm MHz}$. After the rfRAP, the high-offset trap is switched off, leaving a homogeneous offset field of $\simeq 1.2\,{\rm G}$.
 
\section*{Appendix III -- Lifetime of the singlets and triplets}
In order to explain the lifetime $\tau$ of the singlets and triplets observed the experiment (see Fig.~3) we have measured the dependence of $\tau$ on the transverse-lattice and short-lattice depths. In both cases, we find a gradual decrease of $\tau$ with increasing lattice depth. While the decrease in lifetime with short-lattice depth is compatible with the increase in the spontaneous scattering rate of lattice-photons, the dependence on $V_{y,z}$ is much stronger. We explain this behavior by assuming a small spin-dependency of the transverse lattices causing a spin-dependent confinement in the $x$-direction due to the Gaussian shape of the lattice beams.

The red-detuned transverse lattice beams have a waist of $w_{y,z} \simeq 150\,{\rm \mu m}$. The confinement due to their Gaussian profile can be approximated by a harmonic potential with trap frequency 
\begin{equation*}
  \hbar \omega_{x}^{y,z} = \frac{4}{k_{y,z}w_{y,z}}\,  E_r^{y,z} \sqrt{\frac{V_{y,z}}{E_r^{y,z}} - \sqrt{\frac{V_{y,z}}{4 E_r^{y,z}}}}\,,
\end{equation*}
where, $k_{y,z} = 2\pi/\lambda_{y,z}$ is the transverse-lattice wavenumber. If the polarization of the beams has a non-zero circular component with respect to the quantization axis (i.e. the magnetic field), the transverse-lattice depth $V_{y,z}$ will depend on the internal state of the atoms \cite{Deutsch:1998}. Therefore, the harmonic confinement along the $x$-direction will be spin-dependent which translates into a spatially inhomogeneous distribution of $\Delta_m$. Given our experimental parameters and the measured decrease of $\tau$ with $V_{y,z}$, we estimate a relative difference of the transverse-lattice depths for the states $|\!\!\uparrow\rangle$ and $|\!\!\downarrow\rangle$ of $\simeq 0.6\%$. 

The quantization axis in our experiments is mainly oriented along the $x$-direction, thus minimizing spin-dependencies in the transverse lattices. Small angles $\alpha$ between the quantization axis and the $x$-direction, however, are possible. Furthermore, birefringence in the optics passed by the initially linearly polarized lattice beams can yield the required ellipticity of the polarization. For $\alpha \simeq 10°$, an ellipticity of $\epsilon \simeq 0.1$ in the transverse lattice beams leads to the aforementioned spin-dependency of the lattice depth. These values are slightly reduced when also considering the dependency of the spontaneous scattering rate in the lattice on $V_{y,z}$. We finally note, that the dephasing due to a spin-dependent external confinement is stronger for longer-ranged spin-correlations.


\begin{thebibliography}{10}
\bibitem{FermionNote1}%
  \BibitemOpen
  \ \bibinfo {note} {{F}or fermions, the role of the $|t_{0,\pm 1}\rangle$ and
  $|s\rangle$ is reversed. The fact that one atom of a polarized pair $|\rm
  t_{\pm 1}\rangle$ has to occupy the vibrationally excited state is a direct
  manifestation of the Pauli blocking.
   }%
  \bibAnnoteFile{Stop}{FermionNote1}%
\bibitem{ErecoilNote}%
  \BibitemOpen
  \ \bibinfo {note} {{A}ll lattice depths are given in units of the respective
  recoil energy $E_r^{i} = h^2/(2m\lambda_i)$.}%
  \bibAnnoteFile{Stop}{ErecoilNote}%
\end{thebibliography}

%


\end{document}